\documentstyle[11pt]{article}
\catcode`@=11

\ifcase \@ptsize
    q mods for 10 pt
\or 
\or 
\fi

\newtheorem{theorem}{Theorem}[section]
\newtheorem{lemma}[theorem]{Lemma}
\newtheorem{corollary}[theorem]{Corollary}

\newtheorem{definition}[theorem]{Definition}

\newtheorem{fact}{Fact}

\newcommand{\beginproof}{\medskip\noindent{\bf Proof.~}}
\newcommand{\beginproofof}[1]{\medskip\noindent{\bf Proof of #1.~}}
\newcommand{\finishproof}{\hspace{0.2ex}\rule{1ex}{1ex}}
\newenvironment{proof}{\beginproof}{\unskip\nolinebreak\finishproof\par\medskip
}

\title{Lower bounds of quantum 
black-box complexity and degree of approximation polynomials
by influence of Boolean 
variables}
\author{
Yao-Yun Shi
\thanks{This work was supported in part by DARPA and NSF under 
grant number 9627819, and was done when the author
was visiting University of California at Berkeley.}
\\
Computer Science Department\\
Princeton University\\
Princeton, New Jersey 08544\\
E-mail: shiyy@cs.Princeton.EDU}
 
\date{\today}
 
\begin{document}
 
\maketitle

\begin{abstract}
\noindent We prove that, to compute a Boolean function 
$ f : \{ 0, 1\}^N \rightarrow \{ -1, 1\}$ with error probability $\epsilon$, 
any quantum black-box algorithm has to query at least 
$\frac{1 - 2\sqrt{\epsilon}}{2} \rho_f N = \frac{1 - 2\sqrt{\epsilon}}{2} \bar{S}_f$
times, where $\rho_f$ is the average influence 
of variables in $f$, and $\bar{S}_f$ is the average sensitivity.
It's interesting to contrast this
result with the known lower bound of $\Omega (\sqrt{ S_f})$,
where $S_f$ is the sensitivity of $f$.
This lower bound is tight for some functions.
We also show for any polynomial $\tilde{f}$ that approximates $f$ with
error probability $\epsilon$,
$deg(\tilde{f}) \ge \frac{1}{4}( 1 - \frac{ 3 \epsilon}{1 + \epsilon})^2 \rho_f N$.
This bound can be better than previous known lower bound of $\Omega(\sqrt{BS_f})$
for some functions.
Our technique may be of intest itself: we apply Fourier analysis
to functions mapping $\{ 0, 1\}^N$ to unit vectors in a Hilbert space. From this
viewpoint, the state of the quantum computer at step $t$ can be written as
$\sum_{s\in \{0, 1\}^N, |s| \le t} \hat{\phi}_s ( -1 )^ { s \cdot x }$, which 
is handy for lower bound analysis.
\end{abstract}

\section{Introduction}
To compute a Boolean function $f : \{ 0, 1\}^N \rightarrow \{ -1, 1\}$
in the black-box model, the only way the computer can access the input is
to ask an oracle questions like: what is $x_i$.  
The complexity measurement is the number of times the computer asks.
For example, to compute $f ( x_0, x_1, \ldots, x_{N-1} ) = x_0 \vee x_1 \vee \ldots \vee
x_{N-1}$, any classical computer needs to query $\Omega ( N )$ times, while surprisingly, 
there exists a quantum algorithm that queries only $O ( \sqrt{N} )$ times \cite{Gr96}.
The speedup of quantum computers can even be exponential if promise problems are 
considered \cite{DJ92, Si94}.  We study quantum lower bounds 
for error-bounded computation of total Boolean functions only. 

Grover's algorithm was shown to be optimal by
\cite{BBHT96, Za97}, and 
tight bounds for all symmetric Boolean functions,
in particular for some 
familiar ones like $PARITY$, $MAJORITY$,
are shown in \cite{BBCMW98} by
polynomial method. The latter relates
two characterizations of Boolean function complexity, the lowest degree
of approximation polynomials and the block sensitivity,
to lower bounds of quantum black-box complexity. A
Boolean
function $f$ on $N$ Boolean variables 
can be uniquely 
represented as a multi-linear
real polynomial that takes $f(x)$ for $x \in \{ 0, 1\}^N$. 
Let $\tilde{f}$ be a real polynomial that approximates
$f$ and has the least degree.
The block sensitivity of $f$, $BS_f$, is the maximum (over all possible
inputs) number of non-intersecting blocks of 
variables in an input such that flipping all variables in one
block flips the function value. 

\begin{theorem} \cite{BBCMW98}
\label{deg}
To compute a Boolean function $f ( x_0, x_1, \ldots, x_{N-1})$ with error probability
$\epsilon$,,
any quantum computer needs to query at least $\frac{1}{2}deg ( \tilde{f} )$ times.
\end{theorem}

\begin{theorem}\cite{BBCMW98}
To compute a Boolean function $f$ on $N$ variables with error probability $\frac{1}{3}$,
any quantum computer
needs to ask at least $\frac{1}{4}\sqrt{BS_f}$ times.
\end{theorem}

In the latter theorem, replace $BS_f$ by the sensibility $S_f$, 
we get the best known lower bound by of sensitivity.
$S_f$ is just the maximum (over all possible inputs) 
number of bits such that flipping
each flips the function value.

Just like the lowest approximating degree, 
block sensitivity, and sensitivity, 
the influence of variables ($Inf_i(f)$)
and average 
sensitivity ($\bar{S}_f$), introduced by \cite{KKL88}, 
are yet other important characterizations of Boolean function complexity.
One can show that $\sum_i In_i(f) = \bar{S}_f$.
Our main result provides the first known quantum black-box complexity lower 
bound by influence of variables ( or equivalently, by average sensitivity):

\begin{theorem}\label{main}
To compute a Boolean function $f : \{ 0, 1\}^N \rightarrow \{ -1, 1\}$ with error
probability $\epsilon$, any quantum algorithm needs to query at least
$\frac{1 - 2\sqrt{\epsilon}}{2}\rho_f  N = 
\frac{1 - 2\sqrt{\epsilon}}{2} \bar{S}_f$ 
times, where $\rho_f$ is the average influence of variables of $f$.
\end{theorem}

Actually we prove a stronger results:

\begin{theorem}\label{general}
To compute Boolean function $f : \{ 0, 1\}^N \rightarrow \{ -1, 1\}$ with error
probability $\epsilon$, any quantum algorithm needs to query at least
$\frac{1}{2} \cdot \biggl[ 1 - \sqrt[k]{ \frac{ 1 + 2\sqrt{\epsilon}}{2}
         + \frac{1 - 2\sqrt{\epsilon}}{2} \sum_s \hat{f}_s^2 ( 1 - 2 \lambda_s )^k 
         } \biggr] \cdot N$ times, for any positive 
odd integer $k$,
and $\hat{f}_s = E_{x} \Bigl[ f(x) ( -1 )^{ x \cdot s} \Bigr]$.
\end{theorem}

The lowest degree of approximation polynomials for symmetric functions
has a tight characterization \cite{Pa92}. However, for asymmetric
functions, the best general lower bound known is:

\begin{theorem}\cite{NS92} \label{appdeg} For any Boolean function $f$ and its 
approximation $\tilde{f}$,
$deg(\tilde{f}) \ge \sqrt{{BS_f}/6}.$
\end{theorem}

Our lower bound by influence can be better than this bound
for some functions:

\begin{theorem}\label{degree}
For any polynomial $\tilde{f}$
that approximates $f$ with error probability $\epsilon$,
$deg(\tilde{f}) \ge \frac{1}{4} ( 1 - \frac{3\epsilon}{1+\epsilon})^2 \rho_f N$.
\end{theorem}

Note that, Theorem \ref{degree} and Theorem \ref{deg}
together imply a lower bound of quantum black-box complexity
that is asymptotically the same as in Theorem \ref{main}.
However, the former gives a better constant.

Now we turn to our proof ideas. 

For any oracle $x$, let $\phi(x)$ be the state of the quantum computer after
$T$ times of queries. 
 If $f(x) \ne f(y)$, then, in order to distinguish $x$ 
and $y$, $\|\phi(x) -\phi(y)\|^2$ must be large, meaning,
lower bounded by $2 - 4\sqrt{\epsilon}$. If we pick $x$ uniformly random from
$\{ 0, 1\}^N$, and pick a random bit $x_i$, then flip $x_i$ to
get $y$, then we make a hard time for the quantum algorithm: on one hand,
$E = E_{(x, y)}\Bigl[\|\phi(x) -\phi(y)\|^2\Bigr]$ is large (meaning, lower bounded by
$(2 - 4\sqrt{\epsilon}) \rho_f$), on the other hand, this expected value is
upper-bounded by $\frac{4T}{N}$, which then gives us Theorem \ref{main}.

Here is how we get the upper bound of $E$ by $\frac{4T}{N}$.
The key observation of \cite{BBCMW98} is that  
with oracle $x$, after $T$ queries,
the state of the quantum computer can be expressed as:
$$\label{classical}\phi( x ) = \sum_{c} p_c(x) |c\rangle,$$
where $\{c\}$ is a set of orthonormal basis, and $p_c$'s are polynomials
over $x_i$'s of degree at most $T$. 
Since $p_c$'s 
can be represented in Fourier
basis $\{ L_s \}$, where $L_s(x) = ( -1)^{ s \cdot x}$,
we have:
$$\label{equation:rep}\phi( x ) = 
\sum_{ s \in \{ 0, 1\}^N, |s| \le T} \hat{\phi}_s ( -1 )^{ x \cdot s},$$
where $\hat{\phi}_s$ are constant vectors depending on $s$ only.
The number of queries comes in in the maximum possible weight of $s$ with
$\hat{\phi}_s \ne 0$,
since $|s|$ is the degree of polynomial $ ( -1 )^{ x \cdot s}$. 
This representation of $\phi$ is implicit in \cite{FGGS98},
and has a
natural interpretation that will be described later.
We can now write down $E$ explicitly 
in terms of $\hat{\phi}_s$, and then get the desired upper bound by $\frac{4T}{N}$.

For the lower bound of $deg(\tilde{f})$ by average sensitivity, we use the same idea.
We pick the same random pair $( x, y)$, and examine the expected value of
$\| \tilde{f}(x) - \tilde{f}(y) \|^2$ and bound this expectation from both sides --
the upper-bound side by $deg(\tilde{f})$, and the lower-bound side by $\bar{S_f}$.

Section \ref{background} reviews standard notions and facts about Fourier transform and
influence of variables.
Section \ref{blackbox} looks at quantum black-box computation from the viewpoint
of Fourier transform, and section \ref{mainresults} 
provides the proofs, which
is followed by open problems.

Some miscellaneous notations: for any string $s \in \{ 0, 1\}^N$, $|s|$ equals to 
the number of $1$'s in $s$, and $\lambda_s = |s|/N$, $e_i \in \{ 0, 1\}^N$ is the string with
the only $1$ in the $i$'th position, for $0 \le i \le {N-1}$. All 
probability distributions are uniformly random over the corresponding domain. $+$
usually is bitwise $XOR$, and $\cdot$ is usually inner product.

\section{Fourier transform of Boolean functions and influence of variables}
\label{background}
For any $s \in \{ 0, 1\}^N$, let $L_s : \{ 0, 1\}^N \rightarrow \{ -1, 1\}$ be 
$L_s ( x ) = ( -1 )^{ s \cdot x }$ for any $x \in \{ 0, 1\}^N$. $\{ L_s \}$ form
a basis for all functions mapping $\{ 0, 1\}^N$ to ${C}$. 

\begin{lemma}\label{fourier} (Well known)Any $f : \{ 0, 1\}^N \rightarrow {C}$ can 
be uniquely represented as:
$f = \sum_s \hat{f}_s L_s$, where 
$\hat{f}_s = E_x \bigl[ f(x)L_s(x) \bigr]$,
and $max_{ s, \hat{f}_s \ne 0} \{ |s| \} = deg ( f )$, and 
$\sum_s \hat{f}_s \hat{f}_s^* = E_s [ f(s) f^*(s)]$.
\end{lemma}

\begin{definition}\cite{KKL88} The influence of variable $x_i$ on Boolean 
function $f(x_0, x_1, \ldots, x_{N-1})$ is given by:
$Inf_i( f ) = Pr_{x} \Bigl[ f(x) \ne f( x+ e_i) \Bigr],$
the average influence (of variables) is $ \rho_f = E_i \Bigl[ Inf_i ( f ) \Bigr].$
\end{definition}

\begin{lemma}\label{average}\cite{CG88, KKL88} For any Boolean function $f$ on $N$ variables, if
$f = \sum_s \hat{f}_s L_s$, then
$ \rho_f = \sum_s \hat{f}_s^2 \lambda_s.$
\end{lemma} 

\begin{definition}The sensitivity of $f$ on input $x$ is given by
$S_f ( x ) = | \{ i : f( x ) \ne f( x + e_i ) \} |,$
and the average sensitivity is $\bar{S}_f = E_x \Bigl[ S_f ( x ) \Bigr]$.
\end{definition}

It's easy to check:

\begin{lemma} $\bar{S}_f = \rho_f N$.
\end{lemma}

\section{Quantum black-box algorithm from a viewpoint of Fourier transform}
\label{blackbox}
To compute a Boolean function $f$ on $N$ variables,
 a quantum computer uses four sets 
of registers $|i\rangle_I |a\rangle_A |w\rangle_W |r\rangle_R$:  $I$
(index)
has ${\lg{N}}$
 bits, $A$ (answer) and $R$(result) are of one bit each, and $W$ (working)
of a fixed number of bits. The quantum computer is
working in the Hilbert space spanned by the base vectors
$\{|i\rangle_I |a\rangle_A |w\rangle_W |r\rangle_R\}$.An
algorithm $\phi$ that computes $f$ in $T$ queries with error probability at most 
$\epsilon$ can be represented as a sequence
of $2T + 1$ unitary operators on the $H$:
$$\phi: U_0 \rightarrow O \rightarrow U_1 \rightarrow O \rightarrow \ldots 
\rightarrow O \rightarrow U_T,$$
where $U_i$'s are arbitrary unitary transformations defined by the algorithm, 
and $O$'s are the query gates:
$$O : | i \rangle_I | r \rangle_A |w\rangle_W |r\rangle_R
 \rightarrow | i \rangle_I | r + x_i \rangle_A |w\rangle_W |r\rangle_R.$$
The quantum computer starts from the constant vector $| \vec{0} \rangle_{IAWR}$, apply
the sequence of unitary transformations, and then observing the $R$ register yields 
$f(x)$ with error probability smaller than $\epsilon$.

Essentially, $\phi$ defines a function $\{ 0, 1\}^N 
\rightarrow \{ \vec{v} \in H, \| \vec{v} \|^2 = 1 \}$. 
A key observation in \cite{BBCMW98} is,

\begin{lemma}\cite{BBCMW98} On oracle $x$, the final state of the quantum computer can
be written as:
$$\phi ( x ) = \sum_c p_c(x) | c \rangle,$$
where $c$ is taking over all possible configurations of all the registers, and 
$p_c( x )$ is a polynomial mapping $\{ 0, 1\}^N$ to ${C}$ with $deg ( p_c ) \le T$.
\end{lemma}

Therefore, we can think of $\phi$ as a polynomial with coefficients in $H$. Then, any quantum
black-box algorithm defines such a polynomial of degree at most $T$.
Now we  introduce the Fourier transform viewpoint of quantum black-box algorithms. 

\begin{lemma} Let $\phi$ be defined by a quantum black-box algorithm, 
then $\phi$ can be represented as 
$$\phi = \sum_{s, |s| \le T} \hat{\phi}_s L_s,$$
where 
$\hat{\phi}_s = E_x \Bigl[ \phi(x) L_s(x) \Bigr]$, and
$\sum_s \langle\hat{\phi}_s | \hat{\phi}_s\rangle = 1$.
\end{lemma}

Here is a natural interpretation of $\hat{\phi}_s$.
Let's shift the basis for $H$ from $\{ | c \rangle \}$ to that spanned by eigenvectors of the
oracle gates. Denote $F| j \rangle_A$ by $|\zeta_j\rangle_A$, where $F$ is the Hadamard
transform on $Z^m_2$. The following lemma is easy to check.

\begin{lemma}
For a general oracle gate considered in literature, namely, the oracle $x$ is regarded
as  a function $\{ 0, 1\}^n \rightarrow \{ 0, 1\}^m$, and the oracle gate behaves in this way:
$$\label{oldway}O_x | i \rangle_I | j \rangle_A
 = | i \rangle_I  | j + x(i) \rangle_A,$$ 
Then 
$$\label{newway}O_x | i \rangle | \zeta_j\rangle | w \rangle
| r \rangle  = ( -1 )^{ x \cdot e^i_j} | i \rangle | \zeta_j\rangle | w \rangle
| r \rangle,$$
where $e^i_j\in \{ 0, 1\}^{2^nm}$ and has $j$ in the $i$th block, and $0$ in 
other blocks (each block's length is $m$).
\end{lemma}

Now let's see what's going on in a quantum black-box computation, following 
directions of $\{ | i \rangle |\zeta_j\rangle\}$:
We start from the constant vector $|\vec 0\rangle$, make a unitary transformation $U_0$, then
we project into the subspace spanned by some $| i_1 \rangle |\zeta_{j_1}\rangle$, then we make
a query, resulting a sign-flip $( -1 )^{ x \cdot e^{i_1}_{j_1} }$, then we continue our
walk in the same manner: make the second unitary transformation $U_1$, then go into 
subspace of some $| i_2 \rangle |\zeta_{j_2}\rangle$, ask the oracle to flip our sign by
$( -1 )^{ x \cdot e^{i_2}_{j_2} }$, and so on. One good thing about this walk is that every time
we project to a subspace or make a query, for different oracle, the only difference is the 
sign, and this sign is given by
$$ L_x ( e^{i_1}_{j_1} + e^{i_2}_{j_2} + \cdots + e^{i_t}_{j_t}),$$
if we've gone through the path 
$$ p = | i_1 \rangle |\zeta_{j_1}\rangle \rightarrow | i_2 \rangle |\zeta_{j_2}\rangle  
\cdots \rightarrow | i_t \rangle |\zeta_{j_t}\rangle.$$
We use $\alpha_p$ to denote the vector reached by path $p$ with 
oracle $0$, and $\pi_p = e^{i_1}_{j_1} + e^{i_2}_{j_2} + \cdots + e^{i_t}_{j_t}$, then it's easy 
to check:

\begin{lemma}
$\hat{\phi}_s = \sum_{ p, \pi_p = s} \alpha_p.$
\end{lemma}

\section{Main result}
\label{mainresults}
\subsection{Lower bound of quantum black-box complexity}
We're given a Boolean function $f : \{ 0, 1\}^N \rightarrow \{ -1, 1\}$, which
can be represented as $f = \sum_{s \in \{ 0, 1\}^N} \hat{f}_s L_s$.
Let $\phi : \{ 0, 1\}^N \rightarrow H$ be defined by a quantum black-box algorithm that uses
$T$ queries to compute $f$ with error probability bounded by $\epsilon$,
and for any input 
$x \in \{ 0, 1\}^N$, $\phi ( x ) = \sum_{a: |a| \le T} \hat{\phi}_a L_a ( x )$.

Now we take $x$ uniformly random from $\{ 0, 1\}^N$, and take $i_1, i_2, \dots, i_k$ uniformly
and independently random from $\{ 0, 1, \cdots, N-1\}$, for positive 
odd integer $k$. Let
$\vec{i}= e_{i_1} + e_{i_2} + \cdots + e_{i_k}$. 
We would like to bound
$$E = E_{ x, i_1, i_2, \cdots, i_k } \biggl[ 
\Arrowvert \phi ( x ) - \phi ( x + \vec{i} ) \Arrowvert^2
\bigg]$$
from both side -- the lower bound is in in terms of the average sensitivity, 
and the upper bound by $T$, and then get the lower bound for $T$.

To bound $E$ from below, first note that when a pair of input $x, y \in \{ 0, 1\}^N$
have different function value, the corresponding vector $\phi ( x)$ 
and $\phi ( y )$ are far apart:

\begin{fact}
If $f ( x ) \ne f ( y )$, $ \| \phi ( x ) - \phi ( y ) \|^2 \ge 2 - 
4\sqrt{\epsilon}$.
\end{fact}

Therefore the fraction such that $f ( x ) \ne f ( x + \vec{i} )$
bounds $E$ from below. The fraction is given by the following lemma:

\begin{lemma}
$Pr_{x, i_1, i_2, \cdots, i_k} \big[ f ( x ) \ne f ( x + \vec{i} ) \big]
=
\frac{1}{2} - \frac{1}{2} \sum_s \hat{f}_s^2 ( 1 - 2 \lambda_s )^k$.
\end{lemma}

\begin{proof}
\begin{eqnarray}
&&Pr_{x, i_1, i_2, \cdots, i_k} \big[ f ( x ) \ne f ( x + \vec{i} ) \big]
\\
& = & \frac{1}{2} E \big[ 1 - f ( x ) f ( x + \vec{i} ) \big]\\
& = & \frac{1}{2} - \frac{1}{2} \sum_{s_1, s_2} \hat{f}_{s_1} \hat{f}_{s_2} E \big[
       L_{s_1} ( x ) L_{s_2} ( x + \vec{i} ) \big]\\
& = & \frac{1}{2} - \frac{1}{2} \sum_{s_1, s_2} \hat{f}_{s_1} \hat{f}_{s_2}  
      E \big[ ( -1 )^{ x \cdot ( s_1 + s_2 ) + \vec{i} \cdot s_2} \big]\\
& = & \frac{1}{2} - \frac{1}{2} \sum_{s} \hat{f}_s^2 \Big( E_{i\in\{ 0, \ldots, N-1\}}
      \big[ ( -1 )^ {e_i \cdot s} \big] \Big)^k\\
& = & \frac{1}{2} - \frac{1}{2} \sum_s \hat{f}_s^2 ( 1 - 2 \lambda_s )^k
\end{eqnarray}
\end{proof}

When $k = 1$, this probability is exactly the average sensitivity.

Putting the fact and the lemma together we have:

\begin{lemma}
$E \ge ( 2 - 4\sqrt{\epsilon} ) \big( \frac{1}{2} - \frac{1}{2} \sum_s \hat{f}_s^2 ( 1 - 2 \lambda_s )^k 
\big)$
\end{lemma}

Now let's bound $E$ from above, and we'll see how powerful the Fourier
representation of $\phi$ is.

\begin{lemma}
$ E \le 2 - 2( 1 - \lambda_T )^k.$
\end{lemma}

\begin{proof} ( $|a| \le T$.)
\begin{eqnarray}
E & = & E \biggl[ \Arrowvert \sum_{ a, | a | \le T} L_a ( x ) [ 1 - L_a ( \vec{i} ) ] 
        \hat{\phi}_a \Arrowvert^2 \biggr]\\
& = & E \biggl[ \sum_{ a, b} L_{ a + b} ( x ) [ 1 - L_a ( \vec{i} ) ] [ 1 - L_b ( \vec{i} ) ]
        \langle \hat{\phi}_a | \hat{\phi}_b \rangle \biggr]\\
& = & \sum_a E_{ i_1, i_2, \ldots, i_k} \biggl[ ( 1 - L_a ( \vec{i} ) )^2 \biggr] 
      \| \hat{\phi}_a \|^2\\
& = & \sum_a w_a \| \hat{\phi}_a \|^2
\end{eqnarray}

Where $w_a = E_{ i_1, i_2, \ldots, i_k} \bigl[ ( 1 - L_a ( \vec{i} ) )^2 \bigr]$. Now
let's bound $w_a$ by $\lambda_T$, and we think of $a \in \{ -1, 1\}^N$, with
the standard interpretation of $-1$ as $1$ in the original $a$, and $1$ as $0$.

\begin{eqnarray}
w_a & = & 4 Pr \bigl[ a_{i_1} \cdot a_{i_2} \cdots a_{i_k} = -1 \bigr]\\
& = & 4 E \bigl[ \frac{ 1 - a_{i_1} \cdot a_{i_2} \cdots a_{i_k} }{2} \bigr]\\
& = & 2 - 2 ( E_i [ a_i ] )^k\\
& = & 2 - 2 ( 1 - 2 \lambda_a)^k\\
& \le & 2 - 2 ( 1 - 2 \lambda_T)^k
\end{eqnarray}

Note that $\sum_a \| \hat{\phi}_a \|^2 = 1$, therefore,
$$ E \le 2 - 2 ( 1 - 2 \lambda_T )^k.$$
\end{proof}

Now put the above two lemmas together, solve the inequality, then we get 
Theorem \ref{general}. Theorem \ref{main} is obtained by Theorem \ref{general}
and Lemma \ref{average}.

At the time of writing, we don't know if we can  make use of the theorem
for $k>1$. But for $k=1$, any lower bound for sum of influence
implies lower bound for quantum black-box complexity. Here are two examples:

When $f$ is a random function, $\rho_f = \frac{1}{2}$, therefore we have:

\begin{corollary} To compute
 a random function $f$ with error probability $\epsilon$,
the expected number of queries of
any quantum computer is at least $\frac{1 - 2\sqrt{\epsilon}}{4} N$.

\end{corollary}

This matches \cite{Da98} up to a constant factor, 
though slightly worse than the lower 
bound of \cite{Am98}. Since the $PARITY$ function has average sensitivity $1$, we
have
\begin{corollary} To compute 
$PARITY$ with error probability $\epsilon$,
any quantum computer needs to query at least
$\frac{1 - 2\sqrt{\epsilon}}{2} N$ times.
\end{corollary}

This is also a known result \cite{BBCMW98}, \cite{FGGS98}.

\subsection{Degree lower bound of approximating polynomials}
\begin{definition}\cite{NS92}
Let $f : \{ 0, 1\}^N \rightarrow \{0, 1\}$ be a Boolean function, we say 
$\tilde{f} : \{ 0, 1\}^N \rightarrow R$ approximates $f$ with 
error probability $\epsilon$ if for any $x \in \{ 0, 1\}^N$, 
$ | f(x) - \tilde{f}(x) | \le \epsilon.$
\end{definition}

Let $\tilde{f} =  \sum_s \hat{\tilde{f}}_s L_s$, and $d = deg( \tilde{f} )$.
 Define 
$$E' = E_{x, i} \Biggl[ 
| \tilde{f}(x) - \tilde{f}( x + e_i ) |^2 \Biggr].$$
Putting the following two lemmas together we get Theorem \ref{degree}.

\begin{lemma} $E' \ge ( 1- 2\epsilon)^2 \rho_f$.
\end{lemma}

\begin{proof} 
$E' \ge  Pr \bigl[ f(x) \ne f(x + e_i) \bigr] ( 1- 2\epsilon )^2 
= ( 1- 2\epsilon)^2 \rho_f.$
\end{proof}

\begin{lemma}
$E' \le 4 ( 1 + \epsilon )^2 \frac{d}{N}.$
\end{lemma}

\begin{proof}
\begin{eqnarray}
E' & = & E \Biggl[ | \sum_s \hat{\tilde{f}}_s ( 1 - ( -1 )^{ s \cdot e_i } ) L_s(x) |^2
\Biggr]\\
& = & \sum_{s_1, s_2} \hat{\tilde{f}}_{s_1} \hat{\tilde{f}}_{s_2}
E_x \Biggl[ L_{s_1+s_2} ( x )\Biggr] 
E_i\Biggl[( 1 - ( -1 )^{ s_1 \cdot e_i } )( 1- ( -1 )^{ s_2 \cdot e_i } )\Biggr]\\
& = & \sum_s \hat{\tilde{f}}_s^2 E_i\Biggl[( 1 - ( -1 )^{ s \cdot e_i } )^2\Biggr]\\
& = & 4\sum_s \hat{\tilde{f}}_s^2 \lambda_s\\
& \le & \frac{4d}{N}\sum_s \hat{\tilde{f}}_s^2\\
& \le & 4 ( 1+ \epsilon )^2 \frac{d}{N}.
\end{eqnarray}
\end{proof}

Since most functions have large average influence, our
lower bound by influence (Theorem \ref{deg}) 
is better than the bound by block sensitivity (Theorem \ref{appdeg}) for
most functions. An extreme example is $PARITY$.  Here is an 
example of asymmetric function: Let $f ( x_0, x_1, x_2, x_3) =
x_0 ( x_1 - x_2 )^2 + ( 1 - x_0 ) ( x_2 - x_3 )^2$, and 
$f_k$ is obtained by iterate $f$ $k$ times:
$f_k ( X_0, X_1, X_2, X_3) = f ( f_{k-1}(X_0),
f_{k-1}(X_1), f_{k-1}(X_2), f_{k-1}(X_3))$, where $X_i$ is the $i$th block
of $4^{k-1}$ input variables. Then one can show that $BS_{f_k} = 3^k$,
$\bar{S}_{f_k} = 2.5^k > \sqrt{BS_{f_k}}$.
It is conceivable that our lower bound 
is beneficial in proving degree lower bounds of
 asymmetric functions 
when the bound by block sensitivity is not good.
(Although, in our example, $\bar{S}_{f_k}$
is not a tight bound for $\tilde{f}$ either: one can show
that $deg(\tilde{f}) \ge 3^k$.)

\section{Open problems}
So far the lower bound of quantum black-box complexity by
the degree of approximation polynomials implies 
any other (asymptotic) lower bounds.
Is it asymptotically optimal for {\it all} Boolean functions? 

For polynomials that approximates symmetric Boolean functions, 
\cite{Pa92} gives a tight characterization on the lowest degree. However,
we do not know much about the case of asymmetric functions. 
Is there any general tight bound for the lowest degree of 
approximation polynomials?

The relations of block sensitivity and sensitivity and their average cases have been
long-standing open problems \cite{NS92, Ru95, Be96, Ve98}
. Is there any lower bound of quantum black-box complexity
and lower degree of approximation polynomials  by average block 
sensitivity?

\section{Acknowledgment}
The author is grateful to Andy Yao for insightful discussions
and encouragement, 
to Umesh Vazirani, Ashwin Nayak, and Andris Ambainis
 for helpful discussions and warm-hearted hosting, and to Ronald deWolf
for his comments and pointing out some mistakes in the previous version.

\end{document}